\documentclass[a4paper]{jpconf}
\usepackage[latin9]{inputenc}
\usepackage{color}
\usepackage{amsmath}
\usepackage{amssymb}
\usepackage{graphicx}
\usepackage{cite}

\RequirePackage[
   hyperindex,colorlinks,bookmarksnumbered,
   plainpages=true, pdfstartview=FitH,bookmarks=false
]{hyperref}
\hypersetup{linkcolor=blue,urlcolor=blue,citecolor=blue}
\makeatletter

\usepackage{color}
\usepackage{dsfont}
\usepackage{braket}

\newcommand{\rucl}{$\alpha$-RuCl$_3$}

\newcommand{\Ztwo}{\mathbb{Z}_2}
\newcommand{\Uone}{\mathrm{U(1)}}

\begin{document}

\title{Stability of ordered and disordered phases in the Heisenberg-Kitaev model in a magnetic field}

\author{Pedro M C\^onsoli$^1$ and Eric C Andrade$^2$}

\address{$^1$ Institut f\"ur Theoretische Physik and W\"urzburg-Dresden Cluster of Excellence ct.qmat, Technische Universit\"at Dresden, 01062 Dresden, Germany}
\address{$^2$ Instituto de F\'isica de S\~ao Carlos, Universidade de S\~ao Paulo, C.P. 369, S\~ao Carlos, SP, 13560-970, Brazil}

\begin{abstract}
The $S=1/2$ Kitaev honeycomb model has attracted significant attention as an exactly solvable example with a quantum spin liquid ground state. In a properly oriented external magnetic field, chiral Majorana edge modes associated with a quantized thermal Hall conductance emerge, and a distinct spin-disordered phase appears at intermediate field strengths, below the polarized phase.  However, since material realizations of Kitaev magnetism invariably display competing exchange interactions, the stability of these exotic phases with respect to additional couplings is a key issue. Here, we report a 24-site exact diagonalization study of the Heisenberg-Kitaev model in a magnetic field applied in the [001] and [111] directions. By mapping the full phase diagram of the model and contrasting the results to recent nonlinear spin-wave calculations, we show that both methods agree well, thus establishing that quantum corrections substantially modify the classical phase diagram. Furthermore,  we find that, in a [111] field, the intermediate-field spin-disordered phase is remarkably stable to Heisenberg interactions and may potentially end in a novel quantum tricritical point.
\end{abstract}
\date{\today}

\section{Introduction}

Over the past two decades, strong spin-orbit coupling has been recognized as a key ingredient to generate novel states of matter. In the context of magnetic insulators, the interplay of strong spin-orbit coupling and strong electronic correlations provides a route to realize bond-directional interactions between magnetic degrees of freedom. Several instances of this connection arise in Mott insulators constituted of magnetic ions with partially filled $4d$ or $5d$ shells \cite{khaliullin05,witczak-krempa14,rau16,natori16,natori18}.  Among such systems, the so-called Kitaev materials have attracted enormous interest as hosts for the Ising-like bond-dependent interactions that are characteristic of Kitaev's honeycomb model \cite{kitaev06,jackeli09,chaloupka10,chaloupka13,winter17,trebst17,hermanns18,takagi19}.
The Kitaev model describes a system of spins $S=1/2$ located at the vertices of a honeycomb lattice and, remarkably, admits an exact solution over its entire parameter space. In the vicinity of its isotropic point, the ground state is a quantum spin liquid (QSL) \cite{savary17b} in which gapless Majorana fermions hop in a static $\Ztwo$ background. Interestingly, a weak magnetic field opens a gap in the Majorana spectrum and induces a transition to a nonabelian QSL that exhibits chiral Majorana edge modes \cite{kitaev06}.

By now, it is well established that Kitaev interactions are strong in the honeycomb iridates \cite{choi12,singh12} and {\rucl} \cite{plumb14,sears15,banerjee16}, wherein Ir$^{4+}$ and Ru$^{3+}$ ions form effective $j_{\mathrm{eff}}=1/2$ local magnetic moments distributed in stacked honeycomb planes. However, the observation of long-range magnetic order in these materials at low temperatures indicates that other exchange interactions come into play and prevent the stabilization of the Kitaev QSL. In \rucl, the magnetic order can be suppressed by an in-plane magnetic field \cite{sears17,wolter17,janssen19}. For certain field directions, a new phase, appearing between the low-field ordered state and the high-field polarized state, exhibits a half-quantized thermal Hall conductance \cite{kasahara18b,yokoi21} -- a unique signature for a gapped topological QSL \cite{aviv18,ye18,gao19}. Evidence supporting this conclusion was also obtained from magnetocaloric effect experiments \cite{balz2019}, as well as from thermal expansion and magnetostriction measurements \cite{gass2020}.

Numerical studies have uncovered yet another striking feature of the Kitaev model in a magnetic field: For an antiferromagnetic (AF) Kitaev coupling, an intermediate-field spin-disordered phase (IFDP) appears sandwiched between the low-field Kitaev QSL and the polarized phase for a wide range of field orientations \cite{liang18,nasu18,zhu18,gohlke18,hickey19,patel19,hickey21}. An appealing scenario is that this phase is a $\Uone$ QSL with a spinon Fermi surface \cite{liang18,hickey19}, which could explain quantum oscillations observed in the longitudinal thermal conductivity of {\rucl} at low temperatures \cite{czajka21,villadiego21}. However, recent studies suggest that such a $\Uone$ QSL may be unstable at low temperatures \cite{jin21,krueger21}, and further investigations are needed to elucidate the nature of the IFDP.

\section{Model and details of the ED calculation\label{sec:model}}

As a minimal model to describe the physics of Kitaev materials, we consider the nearest-neighbor Heisenberg-Kitaev (HK) Hamiltonian \cite{chaloupka10} in the presence of an external magnetic field
\begin{equation}
\mathcal{H}=J\sum_{\left\langle ij\right\rangle }\textbf{S}_{i}\cdot\textbf{S}_{j}+K\sum_{\left\langle ij\right\rangle _{\gamma}}S_{i}^{\gamma}S_{j}^{\gamma}-\textbf{h}\cdot\sum_{i}\textbf{S}_{i},
\qquad \gamma\in\left\{ x,y,z\right\},
\label{eq:hk}
\end{equation}
where $\gamma$ labels the different links on the honeycomb lattice. For convenience, we use the parametrization $J=A\cos\varphi$ and $K=2A\sin\varphi$, with $A>0$ being an overall energy scale. The field directions are given in the cubic spin basis $\left\{ \hat{\textbf{x}},\hat{\textbf{y}},\hat{\textbf{z}}\right\}$, so that $\textbf{h}\parallel\left[xyz\right]$ reads $\textbf{h}\propto x\hat{\textbf{x}}+y\hat{\textbf{y}}+z\hat{\textbf{z}}$.
At zero field, the Hamiltonian \eqref{eq:hk} has four collinear ordered ground states as a function of $\varphi$ \cite{chaloupka13}. While one of these is ferromagnetic (FM), the other three present AF spin correlations and are thus amenable to spin canting once $h$ is switched on. However, because the Kitaev term breaks spin-rotational symmetry, the response of the system strongly depends on the direction of the field. In particular, spin canting can only occur uniformly if $\textbf{h}$ is perpendicular to one of the cubic axes \cite{janssen16,janssen19}. For all other field orientations, the inhomogeneously canted extensions of the zero-field states compete not only with the polarized state, but also with new symmetry-broken phases that allow more favorable canting mechanisms. This gives rise to complex magnetic orderings and metamagnetic transitions at large $S$ \cite{janssen16,janssen17,price17,consoli20}.

Here, we study the Hamiltonian \eqref{eq:hk} for two representative field directions, $\left[001\right]$ and $\left[111\right]$, by means of two complementary approaches: exact diagonalization (ED) and nonlinear spin-wave theory (NLSWT), as originally presented in Ref.~\cite{consoli20}. Since the latter method does not suffer from finite-size effects, it provides valuable means to investigate the fate of all semiclassical phases.
Our ED calculations were performed on a 24-site cluster that preserves the threefold rotational symmetry of the honeycomb lattice and was subjected to periodic boundary conditions. We exploited the translational invariance of Eq.~\eqref{eq:hk} to operate in a subspace of momentum $\textbf{k}$ and determined ground-state properties via the Lanczos method. To capture phase transitions, we monitored two quantities: The second derivative of the ground-state energy with respect to a control parameter $r$ and the ground-state fidelity $f\left(r\right)=\left|\Braket{\psi\left(r+dr\right)|\psi\left(r\right)}\right|$, which measures the overlap between the ground states at $r$ and $r+dr$. Sharp and smooth peaks in $f\left( r \right)$ are typically associated with first-order and continuous phase transitions, respectively. For the most part, our results were derived by taking $r=h$, in which case we also computed the magnetization per site, $m_{h}$.

\section{ED results and phase diagrams \label{sec:Results}}

\begin{figure}
\centerline{
\includegraphics[width=1.0\textwidth]{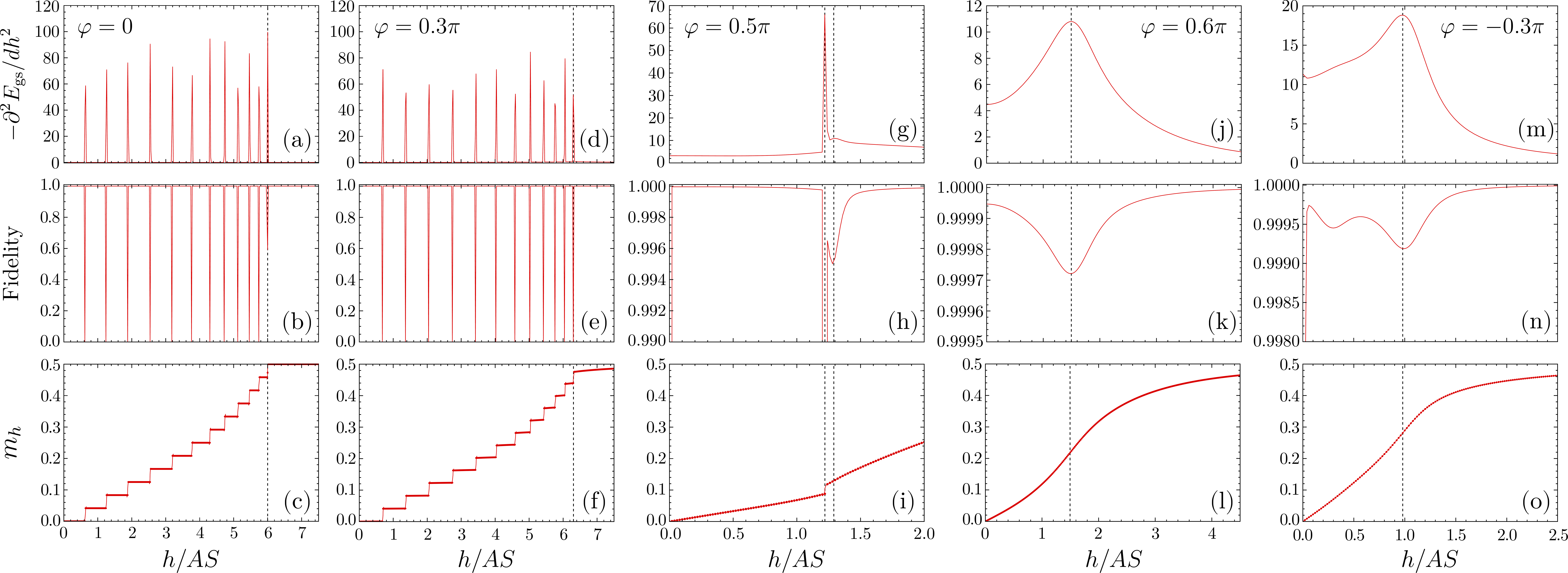}
}
\caption{
\label{fig:edplots001} 
24-site ED results for the HK model in a magnetic field $\textbf{h}\parallel\left[001\right]$ with (a-c) $\varphi=0$, (d-f) $\varphi=0.3\pi$, (g-i) $\varphi=0.5\pi$, (j-l) $\varphi=0.6\pi$, and (m-o) $\varphi=-0.3\pi$. $A$ is a global energy scale related to the coupling constants by $J=A\cos\varphi$ and $K=2A\sin\varphi$ and $S=1/2$. From top to bottom, we show the second derivative of the ground state energy with respect to the magnetic field, the fidelity, and the magnetization per site in the direction of the magnetic field.}
\end{figure}

We begin by analyzing the results for $\textbf{h}\parallel[001]$ and considering the AF Heisenberg point, $\varphi=0$. In Figs. \ref{fig:edplots001}(a-c), we see that both the second derivative of the ground-state energy and the fidelity exhibit a series of sharp peaks as a function of $h$, whereas the magnetization increases in steps. These features simply reflect the finite size of the system and the fact that the total magnetization is a conserved quantity of Heisenberg Hamiltonians. The peak at $h/AS=6$ marks the transition to the polarized phase and survives the thermodynamic limit because the positive curvature of $m_h\left(h\right)$ necessarily leads to a cusp at the critical field.
In Figs. \ref{fig:edplots001}(d-f), we observe a similar behavior for $\varphi=0.3\pi$, but with two important distinctions: (i) The magnetization does not saturate at the last plateau and (ii) the slope of the plateaus now increases with $h$.  Both of these observations are consistent with the fact that the total magnetization ceases to be a conserved quantity of the Hamiltonian when $K\neq0$. Still, the conclusion that all peaks in Figs. \ref{fig:edplots001}(d,e) but the last one are washed out in the thermodynamic limit continues to apply, and we are left with a single continuous phase transition at $h/AS\approx6.49$.

As one moves further away from $\varphi=0$, the magnetization plateaus become steeper and less separated from each other, until they begin to merge near the AF Kitaev point. Fig. \ref{fig:edplots001}(i) shows that the $\varphi=0.5\pi$ magnetization curve has a single discontinuity at $h/AS\approx1.22$, in agreement with Ref. \cite{nasu18}. Yet, as shown in Figs. \ref{fig:edplots001}(g,h), the second derivative of ground-state energy and the fidelity both present a smaller peak at $h/AS\approx1.29$ (the feature close to $h=0$ is likely to be spurious). If one interprets this as a transition to the polarized phase, then the results suggest the existence of an intermediate-field phase. This is precisely the conclusion reached in a previous ED study \cite{patel19}, wherein the intermediate phase was identified as the IFDP.

Past the AF Kitaev point, we find continuous magnetization curves. This is what happens, for instance, at $\varphi=0.6\pi$, see Fig. \ref{fig:edplots001}(l). Since we know that the system is in the zigzag phase at zero field \cite{chaloupka13}, the single smooth peak in Figs. \ref{fig:edplots001}(j,k) must indicate a direct continuous transition to the polarized phase. The situation is not as simple when $\varphi$ belongs to the interval covered by the stripy phase. As illustrated in Figs. \ref{fig:edplots001}(m,n), besides a well-defined peak at $h/AS\approx0.98$, two features appear at lower fields. The sharp peak in the fidelity around $h=0$, in particular, is related to an abrupt growth in the magnitude of the Bragg peaks characterizing the two stripy domains selected by the $\left[001\right]$ field. We did not encounter, on the other hand, signs in the static structure factor capable of explaining the intermediate-field peak in Fig. \ref{fig:edplots001}(n).

\begin{figure}
\begin{centering}
\includegraphics[width=0.48\textwidth]{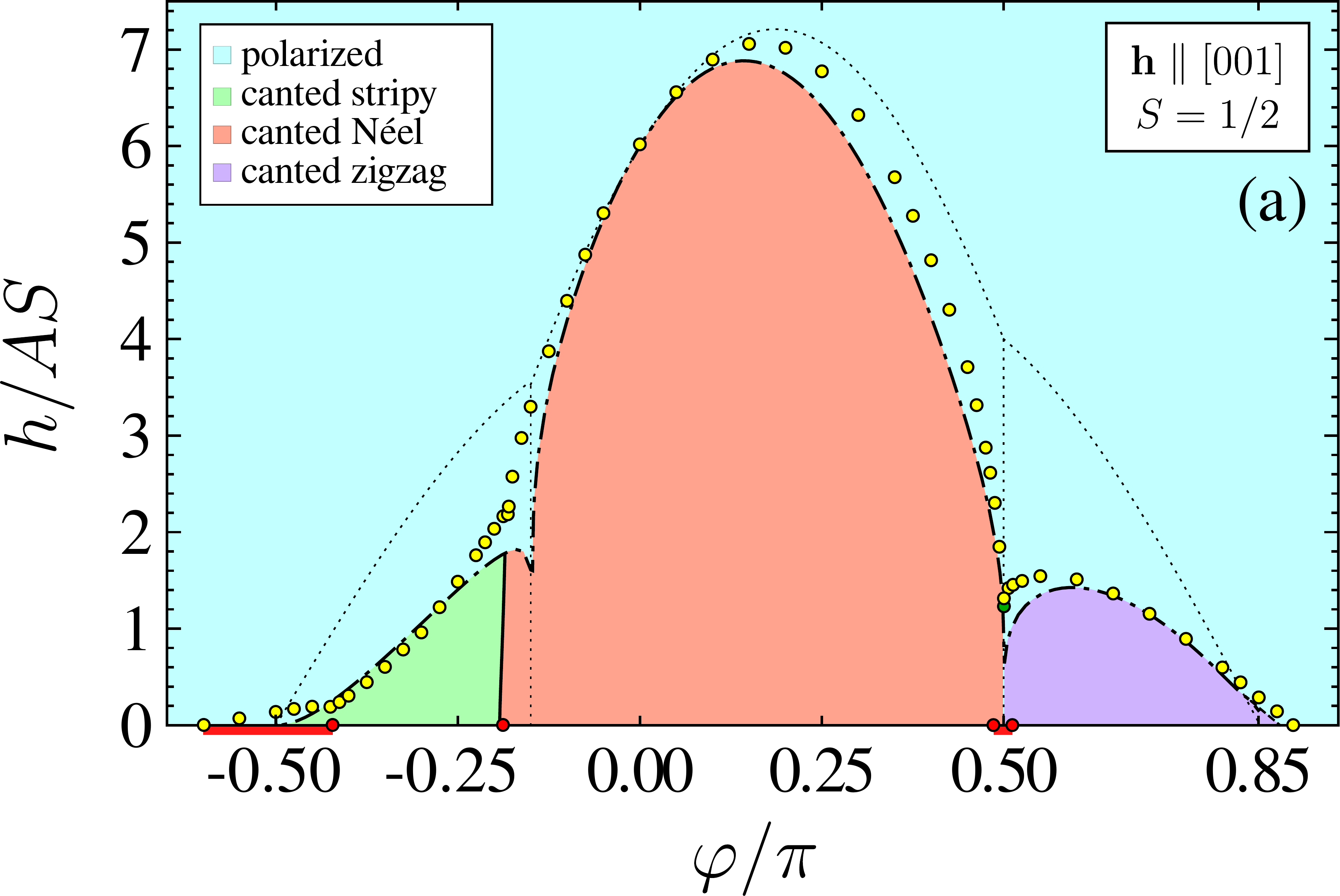}
\hspace{0.04\textwidth}
\includegraphics[width=0.48\textwidth]{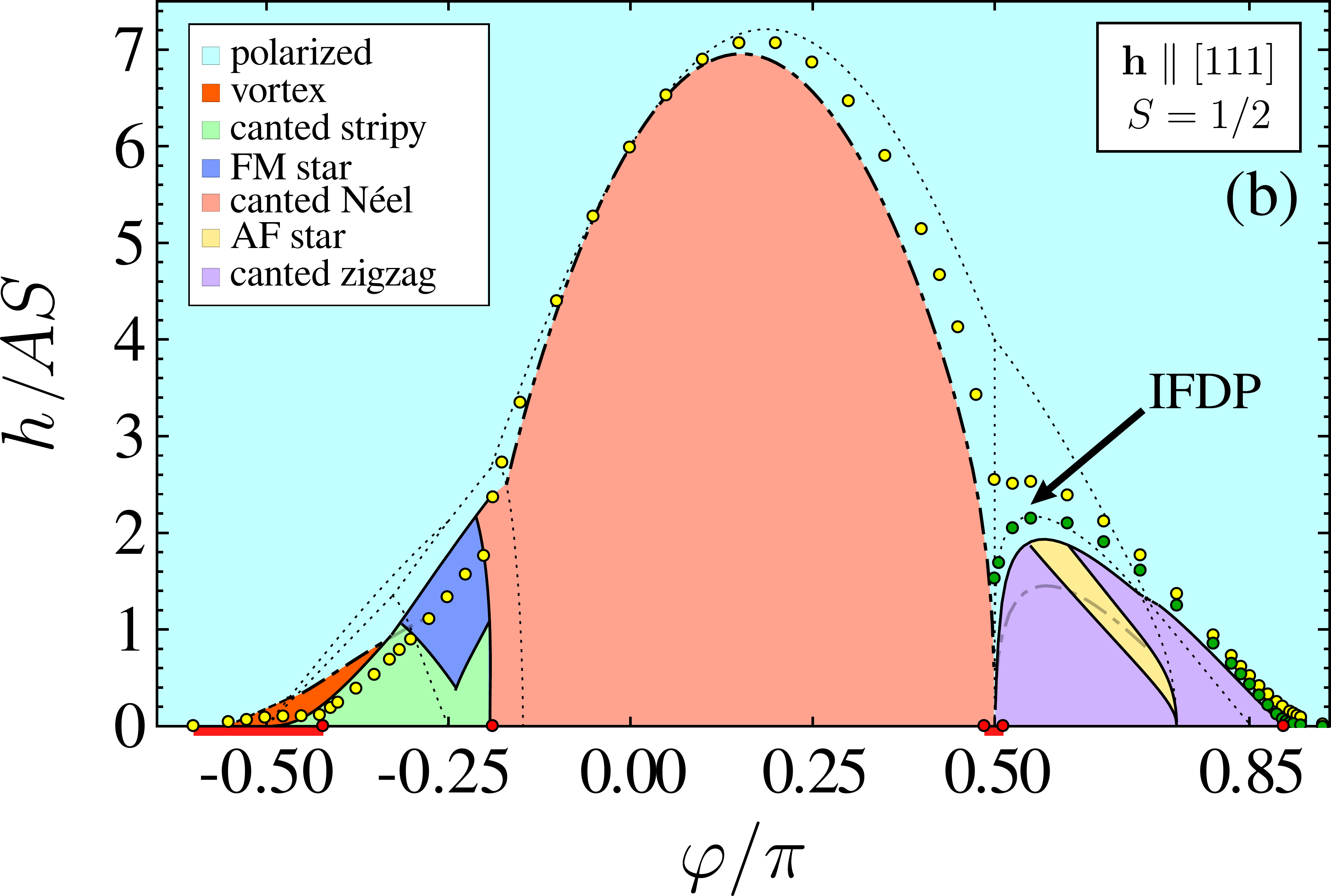}
\end{centering}
\caption{\label{fig:EDpds} $24$-site ED results (dots) superimposed with the $S=1/2$ phase diagrams of the HK model derived within NLSWT
in Ref.~\cite{consoli20} for (a) $\textbf{h}\parallel\left[001\right]$ and (b) $\textbf{h}\parallel\left[111\right]$. Yellow, green, and red dots indicate the last field-induced transition at a given $\varphi$, the lower boundary of the IFDP, and zero-field results, respectively. Red stripes below the horizontal axes highlight the domain of spin-liquid phases at $h=0$, and the dotted lines mark the $S\to\infty$ boundaries.}
\end{figure}

Similar analyses for different $\varphi$ lead to Fig. \ref{fig:EDpds}(a), which compares the ED results with the NLSWT of Ref.~\cite{consoli20}. We do not show data points below the critical fields, however, because secondary features such as those in Figs. \ref{fig:edplots001}(j-k) intermingle with the signatures of phase transitions and render the precise identification of intermediate-field phase boundaries difficult. Overall, we find that both methods are in good agreement for the ordered phases.

\begin{figure}
\centerline{\includegraphics[width=1.0\textwidth]{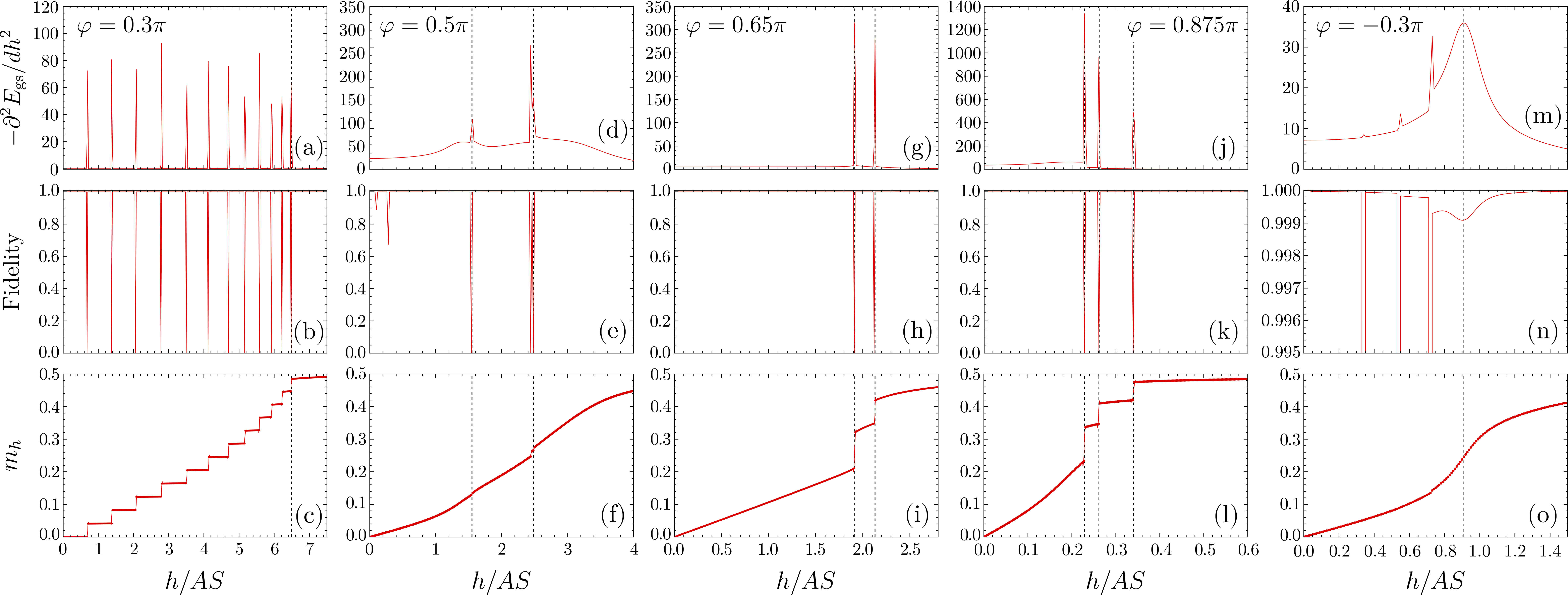}}
\caption{
\label{fig:edplots111}
Same as Fig. \ref{fig:edplots001}, but for a magnetic field $\textbf{h}\parallel\left[111\right]$. (a-c) $\varphi=0.3$, (d-f) $\varphi=0.5\pi$, (g-i) $\varphi=0.65\pi$, (j-l) $\varphi=0.875\pi$, and (m-o) $\varphi=-0.3\pi$.}
\end{figure}

We now discuss the more involved case of $\textbf{h}\parallel\left[111\right]$. For a more in-depth discussion of all semiclassical phases, we refer the reader to Refs.~\cite{janssen16,consoli20}. 
In Figs. \ref{fig:edplots111}(a-c), we see that the data for $\varphi=0.3\pi$ show no qualitative difference with respect to the case of $\textbf{h}\parallel\left[001\right]$. Accordingly, we consider that a single continuous transition occurs between the canted Néel and the polarized phases at the final peak of the second derivative of the ground state and the fidelity.

On the other hand, striking differences appear close to the AF Kitaev point. Figs. \ref{fig:edplots111}(d-f) show two distinct features
at $h/AS\approx1.55$ and $2.48$ delimiting a considerably larger intermediate-field regime than that of Figs. \ref{fig:edplots001}(g-i).
By now, several numerical studies have converged to the conclusion that these peaks represent the boundaries of the IFDP \cite{zhu18,gohlke18,hickey19,ronquillo19,patel19,jiang19,hickey21}. A remarkable property of this phase is its stability to the inclusion of Heisenberg interactions in a $\left[111\right]$ field. While Figs.~\ref{fig:edplots111}(g-i) confirm that the IFDP is present at $\varphi=0.65\pi$, we find that it also extends past $\varphi=0.75\pi$, where the $\left|J\right|>K$. In fact, Figs. \ref{fig:edplots111}(j-l) suggest that it might even be present at $\varphi=0.875\pi$, near the right end of the zigzag phase. The meaning of the third peak appearing in the second derivative of the ground-state energy and in the fidelity remains unclear to us.

Figs.~\ref{fig:edplots111}(m-o) display the results for $\varphi=-0.3\pi$. Once again, we observe a series of features below the smooth peak that signals the transition to the polarized phase. Yet, this time, none of the sharp peaks are associated with the selection of a subset of the three stripy domains, since the $C_{3}^{*}$ symmetry of the Hamiltonian is preserved. This might imply the existence of intermediate-field ordered phases, such as those presented in Refs.~\cite{janssen16,consoli20}. Our preliminary static structure factor results suggest that the vortex phases are absent, as we do not find their corresponding Bragg peaks, but we cannot rule out a transition to the FM star phase.

After assembling results for several other values of $\varphi$, we obtained the phase diagram shown in Fig. \ref{fig:EDpds}(b). As in the [001] case, we refrained from determining the locations of phase boundaries below the upper critical fields. However, we emphasize that our data do not show any evidence for the existence of an ordered phase besides the canted zigzag in the interval $\varphi\in\left[0.5\pi,0.87\pi\right]$. This supports the conclusion from Ref. \cite{consoli20} that the AF vortex is completely suppressed for $S=1/2$, while suggesting that higher-order terms in $1/S$ should account for the disappearance of the remaining slither covered by the AF star. Apart from that, Fig. \ref{fig:EDpds}(b) reveals again a good agreement between our ED and the NLSWT results. We also find consistency with other numerical studies \cite{jiang11,hickey19,jiang19} that focused on reduced portions of the phase diagram.

One novel feature coming from our ED results is the behavior of the upper critical field near the right end of the zigzag phase. The ED results indicate that the two-peak structure from Figs. \ref{fig:edplots111}(g,h) does not give way to a single peak ending at the zero-field transition point $\varphi\approx0.900\pi$. Instead, Fig. \ref{fig:EDpds}(b) suggests that, for $\varphi>0.900\pi$, a $\left[111\right]$ magnetic field could stabilize the IFDP between the zero-field FM state and the high-field polarized phase. While this unlikely scenario is probably linked to finite-size effects, our results raise the question of whether the IFDP actually reaches down to zero field. If so, does it end at a quantum tricritical point shared with the canted zigzag and polarized phases? This intriguing question invites further investigation on the stability of the IFDP and its extent in the phase diagram.

\section{\label{sec:conclusion}Conclusions and outlook}

In this paper, we have studied the stability of various ordered phases of the $S=1/2$ HK model in an external magnetic field. In contrast to spin models with spin-rotational symmetry, where collinear zero-field states typically become simple canted states, the HK model shows, in addition to QSLs, various complex large-unit-cell states in which nonuniform canting may occur.

Our primary goal has been to characterize the effects of quantum fluctuations on the stability of magnetically ordered phases. To this end, we performed a $24$-site ED study and compared our results to those obtained via NLSWT. We were thus able to corroborate the main trend observed in Ref.~\cite{consoli20}, namely that large-unit-cell ordered phases either shrink to smaller portions of the phase diagram or are completely destroyed, while the polarized phase expands. 

The ED study also allowed us to further investigate the fate of the spin-liquid phases of the model. In accordance with previous studies, we find that the FM and AFM Kitaev QSLs have largely asymmetrical responses to a uniform magnetic field and that an IFDP appears in between the low-field AFM QSL and the polarized phase in $\left[001\right]$ and $\left[111\right]$ fields.  Besides this, even though our results do not allow us to discuss the nature of the ground state, they indicate that the IFDP is remarkably stable with respect to the Heisenberg coupling in a $\left[111\right]$ field. We have shown that, in this case, the IFDP survives even for $\left|J\right|\gtrsim K$ and may end at a new quantum tricritical point shared with the canted zigzag and $\left[111\right]$ polarized phases. If confirmed, this can shed more light on the intriguing nature and stability of the IFDP and further highlight the richness of the physics contained in the HK model.

\begin{ack}
We acknowledge discussions with L. Janssen,  H.-H. Tu,  and M. Vojta.  P.M.C.  was supported by FAPESP (Brazil), Grants No. 2017/22133-3 and 2019/02099-0, and from the Deutsche  Forschungsgemeinschaft through SFB 1143 (Project-id No. 247310070). E.C.A. was supported by CNPq (Brazil) and FAPESP (Brazil).
\end{ack}

\section*{References}
\providecommand{\newblock}{}

\end{document}